\documentclass[12pt]{amsart}
\usepackage{amsmath}
\usepackage{latexsym}
\usepackage{amsfonts}
\usepackage{amssymb}
\newcommand{\cal}{\mathcal} 
\newcommand{\bs}{(\Omega,{\cal B}(\Omega))} 
 
\newcommand{\eh}{{\cal E}\,({\cal{H})}}

\newcommand{\pho}{{\cal P}({\cal{H})_1}} 
 
\newcommand{\lh}{{\cal L}({\cal H})} 
 
\newcommand{\hi}{{\cal H}} 
\newcommand{\ip}[2]{\left\langle\,#1\,|\,#2\,\right\rangle} 
\newcommand{\no}[1]{\parallel#1\parallel} 
\newcommand{\ket}[1]{\mid#1\rangle} 
\newcommand{\kb}[2]{|#1\,\rangle\langle\,#2|} 
\newcommand{\fii}{\varphi} 
\newcommand{\R}{\mathbb R} 
\newcommand{\N}{\mathbb N} 
\newcommand{\Z}{\mathbb Z}
\newcommand{\ran}{\operatorname{ran}} 
\newcommand{\bor}[1]{\mathcal{B}(#1)}
\newcommand{\ud}{d}

\newtheorem{lemma}{Lemma}
\newtheorem{corollary}{Corollary}
\newtheorem{proposition}{Proposition}

\theoremstyle{definition}

\newtheorem{example}{Example}

\begin{document}

\title[]{The norm-1-property of a quantum observable}

\author[Heinonen]{Teiko Heinonen}
\address{Teiko Heinonen, Department of Mathematics, 
University of Turku, FIN-20014 Turku, Finland}
\email{teiko.heinonen@utu.fi}
\author[Lahti]{Pekka Lahti}
\address{Pekka Lahti,
Department of Physics, University of Turku, FIN-20014 Turku, Finland}
\email{pekka.lahti@utu.fi}
\author[Pellonp\"a\"a]{Juha-Pekka Pellonp\"a\"a}
\address{Juha-Pekka Pellonp\"a\"a, Department of Physics, University of Turku,
20014 Turku, Finland}
\email{juhpello@utu.fi}
\author[Pulmannova]{Sylvia Pulmannova}
\address{Sylvia Pulmannova,
Mathematical Institute, Slovak Academy of Sciences,
SK--814~73 Bratislava, Slovakia}
\email{pulman@mau.savba.sk}
\author[Ylinen]{Kari Ylinen}
\address{Kari Ylinen, Department of Mathematics, 
University of Turku, FIN-20014 Turku, Finland}
\email{ylinen@utu.fi}

\date{\today}

\begin{abstract}
A  normalized positive operator measure $X\mapsto E(X)$ has the  norm-1-property if $\no{E(X)}=1$ whenever $E(X)\ne O$. This
property reflects the fact that the measurement outcome probabilities
for the values of such observables can be made arbitrary close to one
with suitable state preparations. Some general implications of
the norm-1-property are investigated. As case studies, localization observables, phase
observables, and phase space observables are considered. \\
\noindent
PACS2003 03.65.Ca, 03.65.Ta \\
\noindent
MSC2000 81P15, 81Q10, 81Q99
%noindent
%\bf Keywords:} .
\end{abstract}
\maketitle{}

\section{Introduction}

Spectral measures possess many important properties which have a direct physical meaning 
for the quantum observables represented by such measures. Among them are the following 
properties: 1) the norm of any nonzero operator (projection) in the range of a spectral 
measure is one, 2) the range of a spectral measure is a Boolean $\sigma$-algebra with 
respect to the order structure of operators, 3) any coarse-graining of such a measure is 
a function of that measure. The first property allows one to decide (with probabilistic 
certainty) on the values of the corresponding observable and, for instance, to make the 
variance of such a quantity in a suitable state arbitrarily small. The Boolean structure 
of the range of a spectral measure allows one to combine, in a natural way, statements 
concerning the values (or measurement outcomes) of such observables. Finally, the third 
property is intimately related to the possibility of joint measurability of various 
coarse grainings of such observables. In representing a quantum observable as a 
semispectral measure, i.e. a normalized positive operator measure, one loses, in general, 
the above mentioned properties of spectral measures, and thus also the physical 
interpretation of the relevant measurement context becomes somewhat obscure. In this 
paper we study these properties and their interrelations for semispectral measures and 
we consider their realizations for the approximate localization, phase and the phase 
space observables.

\section{Norm-1-property and $\epsilon$-decidability}

Let $\hi$ be a complex separable Hilbert space
%, inner product $\ip{}{}$ linear in the second argument,
and $\lh$ the set of bounded operators on it. Let $\Omega$ be a nonempty set and 
$\mathcal{A}$ a $\sigma$-algebra of subsets of $\Omega$. Consider a normalized positive 
operator measure $E:\cal A\to\lh$, for short, POM. Such operator measures represent 
physical quantities, observables, of a physical system described by the Hilbert space 
$\hi$. The elements $E(X)$ in the range of $E$, $\ran(E)$, are positive operators bounded 
by the unit operator, that is, $O\leq E(X)\leq I$. Let $\eh$ denote set of operators $A$ 
with $O\leq A\leq I$. They are called effects. Clearly, for any $A\in\eh$, its square 
root $A^{1/2}$ is also an effect with $A\leq A^{1/2}\leq I$. In particular, the square root of 
an effect $A$ is self-adjoint implying $\no{A^{1/2}}^2=\no{A}$. From this equation one 
notices that $\no{A}=1$ if and only if $\no{A^{1/2}}=1$. Also, for any $A\in\eh$, the 
spectrum of $A$, $\sigma(A)$, is a subset of $[0,1]$, and $A$ is a projection operator 
($A^2=A$) if and only if $\sigma(A)\subseteq\{0,1\}$.

We say that a POM $E:\cal A\to\lh$ has the {\em norm-1-property} if the norm of any 
nonzero effect $E(X)$ equals one, that is, $\no{E(X)}=1$, whenever $E(X)\ne O$. Clearly, 
if $E$ is projection valued, that is, each $E(X)$ is a projection operator, then $E$ has 
the norm-1-property.

\begin{lemma}\label{lemma 2}
If $E$ has the norm-1-property, then for any $O\ne E(X)\ne I$, the spectrum of
$E(X)$ contains  $0$ and $1$. %, and, in particular, $E(X)$ is regular.
\end{lemma}
\begin{proof} 
The norm of an effect $E(X)$ is equal to its spectral radius, 
$$
\no{E(X)} \ =\  \sup \{ \lambda \,:\, \lambda\in \sigma(E(X))\}.
$$
Let $X'$ denote the complement of a set $X\subset\Omega$.  If $E$ has the 
norm-1-property, then $\no{E(X)}=1$ as well as $\no{E(X')}=1$ for any $O\ne E(X)\ne I$, 
so that by the closedness of the spectrum, 1 is contained both in $\sigma(E(X))$ and in 
$\sigma(E(X'))$. Since $E(X')=I-E(X)$ and $\sigma(I-E(X))=1-\sigma(E(X))$, one has $0\in 
\sigma(E(X))$.
\end{proof}

We say that a POM $E:\mathcal{A}\to\lh$ has the {\em $\epsilon$-decidability-property}, 
if for each $E(X)\ne O$ and for any $\epsilon>0$ there is a unit vector $\fii$ such that 
$\ip{\fii}{E(X)\fii}\geq 1-\epsilon$.

\begin{proposition}
A POM $E:\cal A\to\lh$ has the norm-1-property if and only if it has
the $\epsilon$-decidability-property.
\end{proposition}

\begin{proof}
For any effect $E(X)$, we have $\no{E(X)}=1$ if and only if $\no{E(X)^{1/2}}=1$. The 
latter equation can be written as
$$
\sup\{\ip{\fii}{E(X)\fii}|\fii\in\hi,\no{\fii}=1\}=1.
$$
\end{proof}

If an observable (POM) $E:\cal A\to\lh$ has the $\epsilon$-decidability-property, then 
for each $X\in\cal A$ for which $E(X)\ne O$ and for each $\epsilon>0$ there is a vector 
state $\fii$ ($\no{\fii}=1$) such that the probability for a measurement of $E$ to lead 
to a result in $X$ in that state $\fii$ is greater than $1-\epsilon$. Since probability 
one and probability almost one are operationally indistinguishable such observables 
resemble sharp observables, that is projection valued observables. The following result, 
known to be valid for sharp observables (spectral measures), exhibits this similarity.

\begin{proposition}\label{variance}
Consider a bounded real POM $E:\mathcal B(\mathbb R)\to\lh$ and 
assume that it has the norm-1-property. Then for each $\epsilon$ there is a vector state 
$\fii\in\hi$ such that $\text{\rm Var}\,(E,\fii)<\epsilon$. \end{proposition}

\begin{proof}
For any $x\in\mathbb R$, $x\in\text{supp}\,(E)$ if and only if for each
$\eta>0$, $E((x-\eta,x+\eta))\ne O$.
Since $E((x-\eta,x+\eta))\ne O$ implies $\no{E((x-\eta,x+\eta))}=1$,
there is a  unit vector $\fii_\eta\in\hi$
such that $\ip{\fii_\eta}{E((x-\eta,x+\eta))\fii_\eta}\geq 1-\eta$. 
Since $\text{supp}\,(E)\subset[-\alpha,\alpha]$ for some $\alpha>0$, we now get
$$
\text{Var}\,(E,\fii_\eta)=\int_{\mathbb R}x^2 d E_{\fii_\eta,\fii_\eta}(x) 
-[\int_{\mathbb R}x\, d E_{\fii_\eta,\fii_\eta}(x)]^2 \leq 15 \eta\alpha^3,
$$ 
which tends to zero with $\eta\to 0$.
\end{proof}

\section{Regular observables and their coarse-grainings}

For any $A\in\eh$ we denote $A':=I-A$ and call it the complement effect of $A$.
If $O\ne A\ne I$, we say that $A$ is {\em regular} if neither $A\leq A'$ nor $A'\leq A$.
A nontrivial effect $A$ ($\ne O,I$) is regular if and only if  $A\not\leq\frac 12I$ and
$A\not\geq \frac 12 I$;
equivalently,  if and only if its spectrum extends
 both below  and above $\frac 12$.
Similarly,
an observable $E:\cal A\to\lh$ is  {\em regular}  if any of its nontrivial
effects $E(X)$ is
regular.

\begin{proposition}\label{Boole1}
If a POM $E$ has the norm-1-property, then it is regular.
\end{proposition}
\begin{proof}
Assume that $E:\cal A\to\lh$ has the norm-1-property.
Then for any $X\in\cal A$, if $O\ne E(X)\ne I$,
we have  $0,1\in\sigma(E(X))$, showing that $E(X)$ is regular.
\end{proof}

The converse statement would be false. As a simple example consider a two-valued POM 
defined as follows: fix a $\lambda\neq \frac{1}{2}$, $0<\lambda <1$, fix also two 
mutually orthogonal unit vectors $\fii,\psi$ and put $A:= \lambda P[\fii]+(1-\lambda 
)P[\psi]$. Then $A$ and its complement $A':=I-A$ constitute a regular POM but $\parallel 
A\parallel = {\mathrm{max}}\{\lambda ,1-\lambda \}< 1$.

Assume  that  a POM  $E:\cal A\to\lh$
has an effect  $E(X)\ne O$ whose norm  is strictly less than 1.
Then 1 is not in the spectrum  of $E(X)$, that is,
0 is not in the spectrum of its complement effect $E(X')$.
Therefore, $E(X')$ is invertible, $\ran(E(X'))=\hi$, and for any
one-dimensional projection operator $P$, the greatest lower bound of
$E(X')$ and $P$ exists and equals
$E(X')\land P =\lambda  P$, where
$\lambda =\no{E(X')^{-1/2}\fii}^{-2}\ne 0$, with
$\fii$ being a unit vector such that $P\fii=\fii$ \cite{BuschGudder}.

Denoting by $\pho$ the set of one-dimensional projections on $\hi$  we may write
any effect $A$  as a join of the weak atoms contained in it, that is,
in the form $A=\lor_{P\in\pho}(A\land P)$ \cite{BuschGudder}. Therefore,
we now have that the set of effects $B$ which are below $A$ and $E(X')$
is different from zero,
that is,
$$
lb\,(A,E(X')) := \{B\in\eh\,:\, B\leq A, B\leq E(X')\}\ne \{O\}.
$$

Consider an arbitrary POM $E:\cal A\to\lh$. The range of $E$ is closed under the mapping 
$E(X)\mapsto E(X)'$. Also the order of effects (as positive operators) may be restricted 
to ${\ran}(E)$. However, the map ${\ran}(E)\ni E(X)\mapsto E(X)'\in{\ran}(E)$ need not be 
an orthocomplementation, since $E(X)\land_{{\ran}(E)}E(X)'$ may fail to exist, and even 
if it does exist, it need not be the null effect $O$. Neither does it need to hold that 
$E(X\cap Y)= E(X)\land_{{\ran}(E)}E(Y)$. In particular, this oddity occurs if $\no{E(X)}<1$ for 
some $E(X)\ne O$. However, if $E$ is regular, then ${\ran}(E)$ is a Boolean lattice with 
respect to the order and the complement restricted to ${\ran}(E)$. The converse is also 
true: if $({\ran}(E),\leq ,')$ is Boolean, then $E$ is regular \cite{DvPu}. In 
particular, in that case we have $E(X)\land_{{\ran}(E)}E(X)'=O$ for any $X\in\cal A$. Any 
(nonzero) lower bound of $E(X)$ and $E(X)'$ (in $\eh$) is necessarily irregular, and as 
such cannot be contained in the range of $E$, which is Boolean.

Consider next two POMs $E$ and $E_1$ defined on the Borel sigma algebras $\bs$ and 
$(\Omega_1,\cal B(\Omega_1))$ of some complete, separable, metric spaces $\Omega$ and 
$\Omega_1$. We say that $E_1$ is a coarse-graining of $E$ if 
$\mathrm{ran}(E_1)\subseteq\mathrm{ran}(E)$. If $E$ is regular, then there is a Borel 
function $f:\Omega\to\Omega_1$ such that $E_1=E^f$, that is, $E_1(Y)=E(f^{-1}(Y))$ for 
all $Y\in\cal B(\Omega_1)$ \cite{LP1}. The converse statement would be false: there are 
irregular observables, e.g. observables with the $\lor$-property (or strong observables) 
such that their coarse grainigs are functions \cite{LP2}.

We close this section with a result concerning finite
coarse-grainings of an observable having the norm-1-property.

\begin{proposition}Let $E:{\cal B}(\Omega) \to \eh$ be an
observable with the  norm-1-property. Let ${\mathcal C}=(A_i)_{i=1}^n$
be a partition of unity in ${\rm ran}(E)$.
Define a mapping $u_{\mathcal C} :\eh \to \eh$ by
$B \mapsto \sum_{i=1}^n A_i^{1/2}BA_i^{1/2}$.
Then for every $1\leq i\leq n$ there is a sequence
$(\psi^i_k)_{k\in\N}$ of unit vectors in $\hi$ such that
$$
\lim_{k\to \infty}\ip{\psi ^i_k}{u_{\mathcal C}(B)\psi ^i_k}
=\lim_{k\to \infty}\ip{\psi^i_k}{A_i^{1/2}BA_i^{1/2}  \psi^i_k}
$$
for all $B\in \eh$.
\end{proposition}

\begin{proof} Owing to the norm-1-property, we can find a sequence of
unit vectors $(\psi ^i_k)_{k\in\N}$
such that $\ip{\psi^i_k}{A_i\psi^i_k} \to 1$, $k\to \infty$.
Since $(A_i)_{i=1}^n$ is a partition of unity, it follows that
$\sum_{j\neq i}\ip{\psi^i_k}{A_j\psi^i_k} \to 0$. From
$$
0\leq \ip{\psi^i_k}{A_j ^{1/2}BA_j ^{1/2}\psi^i_k}
\leq\no{B}\no{A_j ^{1/2}\psi^i_k} \to 0, \  j\neq i,
$$
the desired statement follows.
\end{proof}

\section{Lower bounds for pairs of effects}

The norm-1-property of an observable $E:\cal A\to\lh$ is closely related to the set of 
lower bounds of an effect and its complement. Indeed, if $E$ does not have the 
norm-1-property, then $lb\,(E(X),E(X'))\ne\{O\}$ for some $O\ne E(X)\ne I$. On the other 
hand, if $E$ has the norm-1-property, then $E(X)\land_{\ran(E)}E(X')=O$ for any $X\in\cal 
A$, and any lower bound of $E(X)$ and $E(X')$ (in $\eh$) is necessarily irregular, and as 
such cannot be contained in the range of $E$. These observations call for a further study 
of the set of lower bounds of an effect and its complement.

Let $A,B \in \eh$, and let $A=\lor_{P\in\pho}\lambda(A,P)P$ and
$B=\lor_{P\in\pho}\lambda(B,P)P$.
If $\ran(A^{1/2}) \cap \ran(B^{1/2}) = \{0\}$,
then $A\land B=O$, and  if
$\text{ ran}\,(A^{1/2})\cap \text{ ran}\,(B^{1/2})\ne \{ 0\}$,
then $lb\,(A,B)\ne \{ O\}$.
In the latter case, the greatest lower bound $A\land B$
may or may not exist. In any case, there is always a maximal lower bound.

\begin{proposition}\label{th:1}\cite{MG}  Let $A,B \in \eh$.
There is a maximal $C\in \eh$ such that $C\leq A,B$.
\end{proposition}

\begin{proof}
The set of lower bounds $lb\,(A,B)$ of $A$ and $B$  is a nonempty
partially ordered set in $\eh$.
Let $K\subset \,lb\,(A,B)$ be a chain.
 $K$ is a directed set,
and by indexing its elements by themselves, $K$ becomes an increasing net in $\eh$. 
Applying known results (e.g., \cite[Lemma 1]{T1}, one obtains $C\in \eh$ such that
$$
\lim_ {D\in K} \ip{D\fii}{\fii}=\ip{C\fii}{\fii}
$$
for all $\fii\in \hi$.
It follows that $C\in \,lb\,(A,B)$ and that $D\leq C$ for all $D\in K$.
By Zorn's lemma, $lb\,(A,B)$ has a maximal element.
\end{proof}

\begin{corollary} \label{co:1} Let $A,B\in \eh$.
Then every lower bound of $A,B$ lies under a maximal lower bound.
\end{corollary}
\begin{proof} For every $D_0\in\,lb\,(A,B)$,
put $S(D_0)=\{ D\in\,lb\,(A,B): D_0\leq D\}$.
Then $S(D_0)$ is partially ordered and nonempty, because $D_0\in S(D_0)$.
By the same arguments as above, there is a maximal element in $S(D_0)$.
\end{proof}

For any $A\in\eh$, let $E^A$ denote its spectral measure, so that $A = \int_{0}^{1} 
\lambda \,\ud E^A(\lambda)$. Consider the reduced operators
\begin{eqnarray*}
\widetilde A&:=& A[I- E^A(\{1\})-E^A(\{0\})] =
\int_{0+}^{1-} \lambda  \,\ud E^A(\lambda),\\
\widetilde{I-A}&:=& (I-A)[I - E^A(\{0\})- E^A(\{1\})] =\int_{0+}^{1-} (1-\lambda)\,\ud 
E^A(\lambda),
\end{eqnarray*}
where the spectral projections $E^A(\{0\})$ and $E^A(\{1\})$ are nonzero
exactly when 0 and 1 are eigenvalues of $A$.

\begin{proposition}\label{th:4} \cite{A} The infimum $A\wedge A'$ in $\eh$
exists if and only if the reduced operators $\widetilde A$ and $\widetilde{I-A}$ are 
comparable. In each case, the infimum coincides with the smaller of the above two and is 
equal to
$$
\int_0^1 \min(\lambda ,1-\lambda )\ud E^A(\lambda).
$$
\end{proposition}

\begin{corollary} \label{co:5}
 Let $A\in\eh$. If $0,1$ are not eigenvalues of $A$,
then  $A\land A'$ exists if and only if  $A\leq I-A$ or $A\geq I-A$,
that is, $A$ is irregular.
\end{corollary}

\begin{example}
Let $P_0,P_1,P_2,P_3$ be four mutually orthogonal projections which sum up to
the unit operator, and let $0<\lambda ,\mu <1, \lambda \ne\mu $.
Then $A=0P_0+1P_1+\lambda P_2+\mu P_3$ is an effect
with  $0,1,\lambda ,\mu $ as the eigenvalues.
Then $A$ and $I-A$ are of  norm one, both having 0 and 1 as eigenvalues,
and they constitute a simple observable with the range $\{O,A,I-A,I\}$.
Now $A\land(I-A)$ exists in $\eh$ if and only if the
reduced operators $\widetilde A = \lambda P_2+\mu P_3$
and $\widetilde{I-A}= (1-\lambda )P_2+(1-\mu )P_3$ are comparable.
This is the case exactly when either
$\lambda \leq \frac 12, \mu \leq\frac 12$ or
$\lambda \geq\frac 12, \mu \geq \frac 12$.
In that case $A\land(I-A)$ is the smaller of the two effects
$\widetilde A$ and $\widetilde{I-A}$.
Clearly, $A\land(I-A)$, when it exists, is irregular and is
therefore not contained in the range of the
regular observable in question.
\end{example}

\section{Examples}

In this section properties discussed above are consider for the localization observables, 
the phase observables, and the phase space observables together with their polar and 
cartesian marginal measures.

\subsection{Approximate localization}

Massless relativistic particles are known to be  approximately localizable in the sense that
they admit localization observables $E:\cal B(\mathbb R^3)\to\lh$, 
that is, POMs which are covariant under Euclidean motions and dilations,
 having the norm-1-property for  (nonempty) open sets $U\subset\R^3$, see \cite{cast,Toller,Holevo01}.
For  any nonempty open set $U\subset\R^3$  there is thus a sequence of unit vectors  
$(\psi_n)_{n\in\N}$ such that 
\begin{equation}\label{avoin} 
\lim_{n\to\infty}\ip{\psi_n}{E(U)\psi_n}=1. 
\end{equation} 
Example~\ref{cantor} below shows that
there are (nonnormalized) positive operator measures which do have the
 norm-1-property for open sets but not for all Borel sets. 
Therefore, we shall take a closer look at the norm-1-property. 

\begin{example}\label{cantor} 
Let $C\subset [0,1]$ be a Cantor set with positive Lebesgue measure. It is well-known 
that $C$ is compact and nowhere dense. Define a function $f:\R\to\R$ with 
$f(x)=\frac{1}{2}$ for $x\in C$ and $f(x)=1$ otherwise, and define a (nonnormalized) 
positive operator measure $E:\cal B(\R)\to L^2(\R)$ via the equation $$ 
(E(X)\psi)(x)=\chi_X(x)f(x)\psi(x). $$ For any nonempty open set $U\subset\R$, the 
intersection $U\cap C'$ is an open set with positive Lebesgue measure. Since $E(U\cap 
C')\leq E(U)$, and, by definition, $\no{E(U\cap C')} =1$, it follows that $\no{E(U)} =1$ 
for all open set $U\subset\R$. However, it lacks the norm-1-property, since 
$\no{E(C)}=\frac{1}{2}$.
\end{example}

Let now $\Omega$ be a locally compact second countable topological space
and $\bor{\Omega}$ the Borel $\sigma$-algebra of $\Omega$. 
In this case every  finite Borel measure is a Radon measure. 

\begin{proposition}\label{compact1}
A POM $E:\bor{\Omega}\to\lh$ has the norm-1-property if and only if 
$\no{E(K)}=1$ for all compact sets $K\subset\Omega$ such that $E(K)\neq O$. 
\end{proposition}

\begin{proof} 
Assume that $E$ has the norm-1-property for compact sets $K$ for which $E(K)\ne O$ and 
let $X\in\bor{\Omega}$. If $X$ contains a compact set $K$ such that $E(K)\neq O$, then 
from $E(K)\leq E(X)$ one gets $\no{E(X)}=1$. On the other hand, if $E(K)=O$ for all 
compact sets $K\subset X$, then for any unit vector $\fii\in\hi$,
$$
\ip{\fii}{E(X)\fii}= \sup\{\ip{\fii}{E(K)\fii}\,|\, K\subset X, K \ {\rm compact}\ \}=0,
$$
 showing that also $E(X)=O$. 
\end{proof}

Compact sets are closed. Therefore,  we may replace 'compact' with 'closed' in the previous 
Proposition. This gives us the following formulation, which should be compared with 
equation (\ref{avoin}).

\begin{corollary} If for any open set $U$, $E(U)\neq I$, there is a sequence $(\psi_n)_{n\in\N}$ of unit 
vectors such that
\begin{equation}\label{nolla} 
\lim_{n\to\infty}\ip{\psi_n}{E(U)\psi_n} = 0, 
\end{equation} then $E$ has the norm-1-property. 
\end{corollary}

Condition (\ref{nolla}) means that with a suitable preparation of the state of the system, 
the probability for a measurement result to be in the set $U$ 
can be made arbitrarily small.  It appears  reasonable to expect that an approximate localization 
observable should fulfill also this condition.

Let $G$ be a locally compact second countable group, $H$ a closed and normal subgroup, 
and $\Omega$ the quotient group $G/H$. The Haar measure of $\Omega$ is denoted by 
$\mu_{\Omega}$. Assume that $(U,E)$ is a transitive system of covariance, where $U$ is a 
unitary representation of $G$ in a Hilbert space $\hi$ and $E:\bor{\Omega}\to\lh$ is a 
POM such that $U(g)E(X)U(g)^*=E(g\cdot X)$ for all $g\in G, X\in\bor{\Omega}$.

Lemma \ref{l:haar} below is part of Lemma 3.3 in \cite{Mackey}, and Proposition 
\ref{haar} is a slight modification of Theorem 1 of \cite{Ali}. We find it useful to give 
the proofs of these statements here.

\begin{lemma}\label{l:haar}
Let $\alpha$ be a finite nonzero measure on $\bor{\Omega}$. Then for all 
$X\in\bor{\Omega}$,
\begin{equation}\label{eq:nolla}
\mu_{\Omega}(X)=\frac{1}{\alpha(\Omega)}\int_{\Omega} \alpha(\omega X^{-1})\ud 
\mu_{\Omega}(\omega).
\end{equation} 
\end{lemma}

\begin{proof}
Let $X\in\bor{\Omega}$. The set $\tilde X:=\{(\omega,\eta)\in \Omega\times\Omega | 
\eta^{-1}\omega\in X\}$ is a Borel subset of $\Omega\times\Omega$. 
Clearly, 
$\chi_{\tilde X}(\omega,\eta)=\chi_X(\eta^{-1}\omega)$. Applying the Fubini theorem to 
$\chi_{\tilde X}$ one gets
$$
\int_{\Omega}\left( \int_{\Omega}\chi_X(\eta^{-1}\omega) \ud \mu_{\Omega}(\omega)\right) 
\ud \alpha (\eta) = \int_{\Omega} \left( \int_{\Omega} \chi_X(\eta^{-1}\omega) \ud \alpha 
(\eta)\right) \ud \mu_{\Omega}(\omega).
$$
By the left invariance of the Haar measure $\mu_{\Omega}$, the value of the left side of 
this equality is just $\mu_{\Omega}(X)\alpha(\Omega)$. On the right side we can write 
$\chi_X(\eta^{-1}\omega)=\chi_{\omega X^{-1}}(\eta)$. Now the equation (\ref{eq:nolla}) 
follows.
\end{proof}

\begin{proposition}\label{haar} 
For any $X\in\bor{\Omega}$, $E(X)=O$ if and only if 
$\mu_{\Omega}(X)=0$. 
\end{proposition}

\begin{proof} 
For any $\psi\in\hi$ and $X\in\bor{\Omega}$, 
denote $p_{\psi}(X)=\ip{\psi}{E(X)\psi}$. By lemma \ref{l:haar}, we have 
\begin{equation}\label{eq:haar}
 \mu_{\Omega}(X)=\frac{1}{p_{\psi}(\Omega)}\int_{\Omega} 
p_{\psi}(\omega X^{-1})\ud\mu_{\Omega}(\omega).
 \end{equation}
 Assume that $E(X)=O$. Then 
$p_{\psi}(X)=0$ for all $\psi\in\hi$. Let $\omega\in\Omega$. Because $E$ is covariant, 
there is a $g\in G$ such that $p_{\psi}(\omega X)=p_{U(g)^*\psi}(X)$ for all 
$\psi\in\hi$. Hence $p_{\psi}(\omega X)=0$ and $$ 
\mu_{\Omega}(X^{-1})=\frac{1}{p_{\psi}(\Omega)}\int_{\Omega} p_{\psi}(\omega 
X)\ud\mu_{\Omega}(\omega)=0. $$ Since $\mu_{\Omega}$ is the Haar measure, 
$\mu_{\Omega}(X)=0$ if and only if $\mu_{\Omega}(X^{-1})=0$. 

Assume then that $\mu_{\Omega}(X)=0$. Using equation (\ref{eq:haar}) we see that for any 
$\psi\in\hi$, $p_{\psi}(\omega^{-1} X)=0$ for $\mu_{\Omega}$-allmost all 
$\omega\in\Omega$. Let $\{\varphi_j\}_{j\in\N}$ be an orthonormal basis of $\hi$ and let 
$N_j$ be the set of those $\omega\in\Omega$ for which $p_{\varphi_j}(\omega^{-1} X)$ is 
not zero. Then every $N_j$ as well as $N=\bigcup_{j\in\N}N_j$ are $\mu_{\Omega}$-null 
sets. Assume that $\omega\notin N$. Then for all $k,j\in\N$,
\begin{eqnarray*}
|\ip{\varphi_k}{E(\omega^{-1}X)\varphi_j}| &=& |\ip{E(\omega^{-1}X)^{\frac{1}{2}}\varphi_k}{E(\omega^{-1}X)^{\frac{1}{2}}\varphi_j}| 
\\
& \leq & \no{E(\omega^{-1}X)^{\frac{1}{2}}\varphi_k}\cdot \no{E(\omega^{-1}X)^{\frac{1}{2}}\varphi_j} \\
&=& p_{\varphi_k}(\omega^{-1}X)\cdot p_{\varphi_k}(\omega^{-1}X)=0.
\end{eqnarray*}
From this it follows that for each $j\in\N$, $E(\omega^{-1}X)\fii_j = 0$, and thus 
$E(\omega^{-1}X)=O$. For a fixed $\omega\in N'$ we can take $g\in G$ such that 
$E(\omega^{-1}X)=U(g)E(X)U(g)^*$. This means that $E(X)=O$.
\end{proof}
 
Putting Propositions \ref{compact1} and \ref{haar} together we get following statement.

\begin{proposition} 
A covariant POM $E$ has the norm-1-property if and only if $\no{E(K)}=1$ for any compact set $K$ with positive Haar measure. 
\end{proposition}

We wish to emphasize that it remains an open question if condition (\ref{avoin}) implies 
the norm-1-property for covariant observables.

\subsection{Phase observables}
Phase observables are an important class of physical quantities which can be
represented only in terms of POMs, since there are no phase shift covariant
projection valued measures. Such observables can be characterized in various
equivalent ways, the most direct being as follows.
Let $(\ket n)_{n\in\mathbb N}\subset\hi$ be an orthonormal
basis (number basis) of $\hi$.  Then any sequence of unit vectors
$(\xi_n)_{n\in\mathbb N}\subset\hi$
defines a (phase shift covariant) POM $E:\mathcal B([0,2\pi))\to\lh$
through
$$
E(X) =\sum_{n,m\in\mathbb N}\ip{\xi_n}{\xi_m}\tfrac 1{2\pi} \int_Xe^{i(n-m)x}\,\ud 
x\,\kb nm, \ \ \ X\in \mathcal B([0,2\pi)),
$$
with the (covariance) property
$$
e^{ixN}E(X)e^{-ixN} =E(X\dot+ x), 
$$
where $N=\sum_{n\in\mathbb N}n\kb nn$ and $\dotplus$ denotes addition modulo $2\pi$. 
Conversely, any phase observable is of that form for some sequence of unit vectors 
$(\xi_n)_{n\in\mathbb N}\subset\hi$, see e.g. \cite{CDVLP}. Apart from the trivial phase 
(for which $(\xi_n)_{n\in\N}$ is orthonormal), the simplest among them are the elementary 
phase observables $E_{\mathrm{el}}$, defined by sequences $(\xi_n)_{n\in\mathbb N}$ such 
that $\ip{\xi_n}{\xi_m}=\delta_{nm}$, except for $n=s,m=t,s\ne t$, in which case 
$\ip{\xi_s}{\xi_t}=z$, $0<|z|<1$. Such a phase observable has both regular and irregular 
elements in its range but none of them, except $E_{\mathrm{el}}(X)=I$, has norm one. 
Indeed, the eigenvalues of its effects $E_{\mathrm{el}}(X)$ satisfy $0\leq e_-(X)\leq 
e_0(X)=\frac{\ell(X)}{2\pi} \leq e_+(X)$, with $e_\pm(X) =\frac{\ell(X)}{2\pi}\pm|z|| 
\frac 1{2\pi}\int_Xe^{i(s-t)x}\,\text{dx}|$. Varying $X$ we get both regular and 
irregular effects. But always $\no{E_{\mathrm{el}}(X)}=e_+(X)<1$. Thus 
$\ran(E_{\mathrm{el}})$ is not Boolean and $E_{\mathrm{el}}$ does not have the 
$\epsilon$-decidability property.

The canonical phase observable $E_{\text{can}}:\mathcal B([0,2\pi))\to\lh$ is defined by 
a constant sequence $\xi_n =\xi$ for all $n$. The Hilbert space $L^2([0,2\pi))$ has an 
orthonormal basis $\{e_n\}_{n\in\Z}$, where $e_n$ is the function $x\mapsto \frac 1{\sqrt{2\pi}}e^{inx}$. Let 
$V:\hi\to L^2([0,2\pi)]$ be the isometric linear mapping satisfying $V |n\rangle = e_n$ for all $n\in\mathbb N$. The
Hilbert space $\hi$ can be identified via $V$ with the Hardy subspace $H^2$ of $L^2([0,2\pi)]$. 
 With this identification, $P:=VV^*$ is the orthogonal projection of 
$L^2([0,2\pi)]$ onto $H^2$. Let for $X\in\mathcal B([0,2\pi))$,  $M_{\chi_X}$ be the multiplication operator acting on 
$L^2([0,2\pi)]$, $M_{\chi_X}f=\chi_Xf$. It is easy to see that
$$
E_{\text{can}}(X)=V^*M_{\chi_X}V=V^*PM_{\chi_X}V.
$$
The spectra of the operators $E_{\text{can}}(X)$ and $PM_{\chi_X}$ are therefore the 
same. On the other hand, by the Hartman-Wintner theorem \cite[p. 183]{Douglas72} the spectrum of 
the Toeplitz operator $PM_{\chi_X}$ is the closed interval 
$[\text{ess\,inf}\,\chi_X,\text{ess\,sup}\,\chi_X]$. Hence the following proposition is 
obtained.

\begin{proposition}\label{toepliz}
For any $X\in\mathcal B([0,2\pi))$ of nonzero Lebesgue measure the norm $||E_{\text{can}}(X)||=1$.
Moreover, if also the complement set $X'$ has nonzero measure, then the spectrum of $E_{\text{can}}(X)$
is the whole interval $[0,1]$.
\end{proposition}

The norm-1-property of $E_{\text{can}}$ implies that $E_{\text{can}}$ is regular. While 
$0,1\in\sigma(E_{\text{can}}(X))$ for any $O\ne E_{\text{can}}(X)\ne I$, it is well known 
that they are not eigenvalues of $E_{\text{can}}(X)$, see e.g. \cite[p. 5929]{BLPY}. 
Therefore $E_{\text{can}}(X)\wedge E_{\text{can}}(X')$ does not exist in $\eh$. It 
follows that there exist at least two incomparable lower bounds of $E_{\text{can}}(X)$ 
and $E_{\text{can}}(X')$ in $\eh$. Apart from that, $(\ran(E_{\text{can}}),\leq,')$
 is Boolean and $E_{\text{can}}$ has the $\epsilon$-decidability property.

The norm-1-property of the canonical phase observable has been obtained by elementary methods already in
\cite{BLPY}. These methods were needed to study also some properties of the phase space observables,
see below.  For the present purpose we find it useful to give an independent proof for Proposition~\ref{toepliz}.

\subsection{The phase space observable $A_{\ket 0}$} %begin{fact}{\rm
As another physically relevant example, consider the 2-dimen\-sio\-nal phase space 
observable $A_{\ket 0}$ generated by the ground state $\ket 0$ of the number operator 
$N=\sum_{n=0}^\infty n\kb nn$. As is well known, the phase space observable $A_{\ket 0}$ 
has the structure
\begin{equation*}\label{phasespace}
A_{\ket 0}(Z) = \tfrac 1\pi\int_Z\kb zz\, \ud \lambda(z), \ \ \ Z\in\mathcal B(\mathbb 
C),
\end{equation*}
where $\ket z=e^{-|z|^2/2}\sum_{n=0}^\infty \frac{z^n}{\sqrt{n!}}\ket n$
is a coherent state (for each $z\in\mathbb C$)
and  $\lambda:\mathcal B(\mathbb C)\to[0,\infty]$
the two-dimensional Lebesgue measure.
For any $Z\in\mathcal B(\mathbb C)$ of finite measure one has
$A_{\ket 0}(Z)\leq \tfrac{\lambda(Z)}{\pi}I$,
showing that there are effects $A_{\ket 0}(Z)$ with norm strictly less
than one, even less than $\tfrac 12$.
Therefore, the phase space
observable $A_{\ket 0}$ does not have the norm-1-property and is irregular.
Its range is not Boolean.

\subsubsection{Polar coordinate marginal measures}
Using the polar decomposition of complex numbers ($z=re^{i\theta}$),
consider  a set of the form $Z=[0,r)\times[0,2\pi)$,
so that $\lambda(Z)=\pi r^2$ and thus $\no{A_{\ket 0}(Z)}\leq r^2$.
This shows that not only the phase space observable $A_{\ket 0}$
but also its number margin
$$
\mathcal B([0,\infty))\ni R\mapsto A_{\ket 0}(R\times[0,2\pi))
=:A_{\ket 0}^r(R)\in\lh
$$
fails to  have the norm-1-property and is irregular.
On the other hand, if we consider sets of the form
$Z=[0,\infty)\times\Theta$, with $\Theta\in\mathcal B([0,2\pi))$,
the estimate $\ip{\fii}{A_{\ket 0}(Z)\fii}\leq \lambda(Z)/\pi,
\fii\in\hi,\no{\fii}=1$,
does not help to bound the norm of the effect $A_{\ket 0}(Z)$.
However, it can be shown  \cite{BLPY} that the angle margin of $A_{\ket 0}$,
that is, the POM
$$
\mathcal B([0,2\pi))\ni \Theta\mapsto A_{\ket 0}([0,\infty))\times\Theta)
=:A_{\ket 0}^\theta(\Theta)\in\lh
$$
has the norm-1-property:
for any $\Theta\in \mathcal B([0,2\pi))$ of nonzero Lebesgue measure,
$$
\no{A_{\ket 0}^\theta(\Theta)}=1.
$$
Therefore, $A_{\ket 0}^\theta$ is regular and it has the $\epsilon$-decidability property.

\subsubsection{Cartesian marginal measures}
Consider next the norm properties of the Cartesian marginal
(with respect to $z=x+iy$) measures
\begin{eqnarray*}
\mathcal B(\mathbb R)\ni X\mapsto A_{\ket 0}(X\times\mathbb R)
&=:&A_{\ket 0}^x(X) \in\lh,\\
\mathcal B(\mathbb R)\ni Y\mapsto A_{\ket 0}(\mathbb R\times Y)
&=:&A_{\ket 0}^y(Y)\in\lh.
\end{eqnarray*}
This is most readily done by using the $L^2(\mathbb R)$-realization of $A_{\ket 0}$ 
(obtained via the isometry $\hi\ni\ket n\mapsto f_n\in L^2(\mathbb R)$, where $f_n$ is 
the $n$-th Hermite function). In that representation the marginal measures are identified 
respectively as unsharp position and unsharp momentum with the effects $A_{\ket 
0}^x(X)\equiv (|f_0|^2*\chi_X)(\frac 1{\sqrt 2}Q)$ and $A_{\ket 0}^y(Y)\equiv (|\hat 
f_0|^2*\chi_X)(\frac 1{\sqrt 2}P)$, where $Q$ and $P$ are the usual position and momentum 
operators and $\hat f_0$ is the Fourier transform of $f_0$ \cite{Davies}. (We recall that 
it is customary to use the coordinates $q=x/\sqrt 2, p=y/\sqrt 2$ for position and 
momentum observables.) Using the spectral calculus one gets the norm estimates
\begin{eqnarray*}
\no{A_{\ket 0}^x(X)} &=& \no{(|f_0|^2*\chi_X)
(\tfrac 1{\sqrt 2}Q)} \ \leq \
\sup_{x\in\mathbb R}|(|f_0|^2*\chi_X)(\tfrac 1{\sqrt 2}x)|,\\
\no{A_{\ket 0}^y(Y)} &=& \no{(|\hat f_0|^2*\chi_Y)
(\tfrac 1{\sqrt 2}P)}\ \leq \
\sup_{y\in\mathbb R}|(|\hat f_0|^2*\chi_Y)(\tfrac 1{\sqrt 2}y)|.
\end{eqnarray*}
This shows that, e.g.
$\no{A_{\ket 0}^x((-\epsilon,\epsilon))} \leq 2\epsilon/\sqrt\pi$,
which is less than $\frac 12$ whenever $\epsilon<\sqrt\pi/4$.
An immediate computation also shows  that for any bounded $X\in\mathbb R$,
$\sup_{x\in\mathbb R}|(|f_0|^2*\chi_X)(\tfrac 1{\sqrt 2}x)|<1$
and thus $\no{A_{\ket 0}^x(X)}<1$,
whereas for complements of bounded sets $X$ one gets
$\no{A_{\ket 0}^x(X')}=1$.
Finally, we observe that for any regular effect $A_{\ket 0}^x(X)$,
the meet $A_{\ket 0}^x(X)\land A_{\ket 0}^x(X)'$
does not exist in $\eh$.
On the other hand, if $A_{\ket 0}^x(X)$ is irregular, then
$\{O,A_{\ket 0}^x(X),A_{\ket 0}^x(X)',I\}$ is non-Boolean.

\subsubsection{Two-photon coherent  state probability measures}

The fact that the angle margin $A_{\ket 0}^\theta$ has the norm-1-property means that for 
any $\Theta\in \mathcal B([0,2\pi))$ (of nonzero measure) there is a sequence of unit 
vectors $\fii_n\in\hi$ such that the probabilities $\ip{\fii_n}{A_{\ket 
0}^\theta(\Theta)\fii_n}$ tend to one with growing $n$. In fact, choosing a coherent 
state $\ket\alpha$, $\alpha\in\mathbb C$, such that $\arg\alpha\in \Theta$ is a Lebesgue 
point of $\Theta$, then $\lim_{|\alpha|\to\infty}\ip{\alpha}{A_{\ket 
0}^\theta(\Theta)|\alpha}=1$, see \cite{BLPY}. On the other hand, our investigations of 
Cartesian marginal measures show that for no bounded $X\in\bor{\R}$, is there a sequence 
$(\fii_n)$ of unit vectors for which the probabilities $\ip{\fii_k}{A_{\ket 
0}^x(X)\fii_k}$ would tend to one. We state this obvious fact since one might expect 
that, for instance, squeezing the vacuum state $\ket 0$, $S(r)\ket 0 = 
e^{ra^2-ra^*{}^2}\ket 0$, and rotating and displacing it appropriately, the probability 
$\ip{0}{S(r)^*A_{\ket 0}^x(X)S(r)|0}$ would tend to one (with growing squeeze parameter 
$r$). That this is not the case is seen directly from these probabilities. Instead of 
considering coherent and squeezed states we study directly the more general case of 
two-photon coherent states \cite{Yuen}.

Let $|\beta;\mu,\nu\rangle$, $\beta$, $\mu$, $\nu\in\mathbb C$, $|\mu|^2-|\nu|^2=1$,
be a two-photon coherent state (TCS), that is, it satisfies the following
eigenvalue equation
$$
(\mu a+\nu a^*)|\beta;\mu,\nu\rangle=\beta|\beta;\mu,\nu\rangle.
$$
An arbitrary TCS $|\beta;\mu,\nu\rangle$ 
can be written in the form 
$$
e^{-i\theta/2}R(\theta)D(z)S(\epsilon)\ket0
$$
where $R(\theta):=e^{i\theta N}$, $\theta\in[0,2\pi)$,
$D(z):=e^{za^*-\overline{z}a}$, $z\in\mathbb C$,
$S(\epsilon):=e^{\frac{1}{2}\overline{\epsilon}a^2-\frac{1}{2}\epsilon{a^*}^2}$,
$\epsilon\in\mathbb C$, are the rotation, displacement, and squeezing operators,
respectively.
Note that $|\mu|^2-|\nu|^2=1$ implies that $|\mu|\ge1$ and
$|\nu/\mu|\in[0,1)$. We go on to determine  the density of the probability
measure
$Z\mapsto\langle\beta;\mu,\nu|A_{|0\rangle}(Z)|\beta;\mu,\nu\rangle$.

Let $\mu$, $\nu$ and $\beta$ be fixed. From  \cite[Eq. (3.20)]{Yuen} one gets
$$
\langle z|\beta;\mu,\nu\rangle=\frac{1}{\sqrt{\mu}}
\exp\left(-\frac{1}{2}|z|^2-\frac{1}{2}|\beta|^2-\frac{\nu}
{2\mu}\overline{z}^2+
\frac{\overline{\nu}}{2\mu}\beta^2+\frac{1}{\mu}\overline{z}\beta\right)
$$
for all $z$, $\beta\in\mathbb C$.
Denote $\gamma=\overline{\mu}\beta-\nu\overline{\beta}$. Then $\beta=\gamma\mu+\overline{\gamma}\nu$.
Defining $z':=z-\gamma$ one gets
$$
\langle z|\beta;\mu,\nu\rangle=\frac{1}{\sqrt{\mu}}
\exp\left(-\frac{1}{2}|z'|^2-\frac{\nu}{2\mu}\overline{z'}^2+
\frac{1}{2}\overline{z'}\gamma-\frac{1}{2}z'\overline{\gamma}\right)
$$
and, thus,
$$
\left|\langle z|\beta;\mu,\nu\rangle\right|^2
=\frac{1}{|\mu|}
\exp\left[-|z'|^2-\mathrm{Re}\left(\frac{\nu}{\mu}
\overline{z'}^2\right)\right]
=\left|\langle z'|0;\mu,\nu\rangle\right|^2.
$$
Using the above equation one easily sees that
$$
\langle\beta;\mu,\nu|A_{|0\rangle}(Z)|\beta;\mu,\nu\rangle=
\frac{1}{\pi}\int_Z\left|\langle z|\beta;\mu,\nu\rangle\right|^2\ud \lambda(z)=
\langle0;\mu,\nu|A_{|0\rangle}(Z-\gamma)|0;\mu,\nu\rangle
$$
for all $Z\in\mathcal B(\mathbb C)$.

Next we calculate the Cartesian margins of the probability measure
$$
Z\mapsto\langle\beta;\mu,\nu|A_{|0\rangle}(Z)|\beta;\mu,\nu\rangle.
$$
Now for all $X$, $Y\in\mathcal B(\mathbb R)$,
\begin{eqnarray*}
\langle\beta;\mu,\nu|A_{|0\rangle}^x(X)|\beta;\mu,\nu\rangle&=&
\frac{1}{\sqrt{\pi}|\mu|\sqrt{1-\mathrm{Re}(\nu/\mu)}}
\int_{X-\mathrm{Re}\,\gamma}
\exp\left\{-{x^2}{}\left[\frac{1-|\nu/\mu|^2}
{1-\mathrm{Re}(\nu/\mu)}\right]\right\}\ud x,\\
\langle\beta;\mu,\nu|A_{|0\rangle}^y(Y)|\beta;\mu,\nu\rangle&=&
\frac{1}{\sqrt{\pi}|\mu|\sqrt{1+\mathrm{Re}(\nu/\mu)}} \int_{Y-\mathrm{Im}\,\gamma} 
\exp\left\{-{y^2}{}\left[\frac{1-|\nu/\mu|^2} {1+\mathrm{Re}(\nu/\mu)}\right]\right\}\ud 
y,
\end{eqnarray*}
so that the variances of these probability measures are
\begin{eqnarray*}
\text{\rm Var}\,(A_{|0\rangle}^x,\ket{\beta;\mu,\nu}) &=&
\frac{1}{2}\cdot\frac{1-\mathrm{Re}(\nu/\mu)}{1-|\nu/\mu|^2}\ge\frac{1}{2},\\
\text{\rm Var}\,(A_{|0\rangle}^y,\ket{\beta;\mu,\nu})&=&
\frac{1}{2}\cdot\frac{1+\mathrm{Re}(\nu/\mu)}{1-|\nu/\mu|^2}\ge\frac{1}{2}.
\end{eqnarray*}
Thus, there is no sequence of TCS:s for which the limit measure of corresponding 
cartesian marginal probability measures is concentrated on a point. The uncertainty 
product, the product of the variances of the marginal measures is
$$
\text{\rm Var}\,(A_{|0\rangle}^x,\ket{\beta;\mu,\nu})\cdot
\text{\rm Var}\,(A_{|0\rangle}^y,\ket{\beta;\mu,\nu})=
\frac{1-\left(\mathrm{Re}(\nu/\mu)\right)^2}
{4\left(1-|\nu/\mu|^2\right)^2}\ge\frac{1}{4}
$$
and the lower bound is approached if and only if $\nu=0$ ($|\mu|=1$), that is, the 
corresponding TCS is a coherent state (up to a physically irrelevant phase factor).

When we denote $\beta\equiv se^{i\varphi}$, $s\in[0,\infty)$, $\varphi\in[0,2\pi)$, 
$\theta_\mu:=\arg\mu$, $\theta_\nu:=\arg\nu$, the probability density of the angle margin 
$A_{\ket 0}^\theta$ in the state $|\beta;\mu,\nu\rangle$ gets the form
\begin{eqnarray*}
&&g_{|\beta;\mu,\nu\rangle}(\theta):=\frac{1}{\pi}\int_0^\infty\left|\left\langle re^{i\theta}|
se^{i\varphi};\mu,\nu\right\rangle\right|^2
r\ud r\\
&&=\frac{1}{|\mu|}\exp\left\{-\left[1-\left|\frac{\nu}{\mu}\right|
\cos(2\varphi-\theta_\mu-\theta_\nu)\right]s^2\right\}
\times
\Bigg\{
\frac{1}{2\pi[1+|\nu/\mu|\cos(2\theta-\theta_\mu+\theta_\nu)]}\\
&&+\frac{s\cos(\theta+\theta_\mu-\varphi) 
\exp\left\{s^2\cos^2(\theta+\theta_\mu-\varphi)/ 
\left[|\mu|^2+|\nu\mu|\cos(2\theta-\theta_\mu+\theta_\nu)\right]\right\}} 
{2\sqrt{\pi}|\mu|\left[1+|\nu/\mu|\cos(2\theta-\theta_\mu+\theta_\nu)
\right]^{3/2}}\\
&&\times\left\{1+{\rm erf}\left[\frac{s\cos(\theta+\theta_\mu-\varphi)}{|\mu|
\sqrt{1+|\nu/\mu|\cos(2\theta-\theta_\mu+\theta_\nu)}}\right]\right\} \Bigg\}.
\end{eqnarray*}
When $|\beta;\mu,\nu\rangle$ is a coherent state $|\beta\rangle$
($\mu=1$ and $\nu=0$) then
$$
g_{|\beta\rangle}(\theta)\to\delta_{2\pi}(\theta-\varphi)
$$
when $s\to\infty$.
Also if $s$ is fixed and $\varphi=(\theta_\mu+\theta_\nu)/2$ then
if $|\nu|\to\infty$
$$
g_{|\beta;\mu,\nu\rangle}(\theta)\sim
\frac{1}{2\pi}\frac{1}{|\mu|+|\nu|\cos(2\theta-\theta_\mu+\theta_\nu)}\to
\delta_\pi(\theta-\theta_\mu/2+\theta_\nu/2+\pi/2)/2
$$
($\pi$-periodic Dirac delta). In particular, this holds for a squeezed
and rotated vacuum ($s=0$).

\end{document}